\documentclass[conference]{IEEEtran}
\IEEEoverridecommandlockouts
\usepackage{cite}
\usepackage{mathtools}
\usepackage{algorithmic}
\usepackage{graphicx}
\usepackage{textcomp}
\usepackage[colorlinks]{hyperref}
\usepackage[nameinlink]{cleveref}

\def\BibTeX{{\rm B\kern-.05em{\sc i\kern-.025em b}\kern-.08em
    T\kern-.1667em\lower.7ex\hbox{E}\kern-.125emX}}

\usepackage{microtype}
\usepackage{booktabs,makecell,tabularx}

\newcolumntype{C}[1]{>{\centering\arraybackslash}p{#1}}
\newcolumntype{L}{>{\raggedright\arraybackslash}X}
\newcommand{\specialcell}[2][l]{%
  \begin{tabular}[#1]{@{}l@{}}#2\end{tabular}}
\usepackage{siunitx}
\usepackage{etoolbox}
\newrobustcmd{\B}{\bfseries}  

\begin{document}

\title{Towards the Effects of Alignment Edits on the Quality of Experience of 360 Videos\\
\thanks{This work is partially supported by FAPESP (Fundação de Amparo à Pesquisa do Estado de São Paulo), Grant 18/23086-1, by CAPES (Coordination for the Improvement of Higher Education Personnel), and by the University of Brasília.}
}

\author{
\IEEEauthorblockN{1\textsuperscript{th} Lucas Althoff}
\IEEEauthorblockA{\textit{CNRS, Xlim UMR 7252} \\
\textit{Université of Poitiers}\\
Poitiers, France \\
0000-0002-3387-9686}
\and
\IEEEauthorblockN{2\textsuperscript{th} Alessandro Rodrigues}
\IEEEauthorblockN{3\textsuperscript{th} Mylène C. Q. Farias}
\IEEEauthorblockA{\textit{Computer Science Department} \\
\textit{Universidade de Brasília}\\
Brasília, Brazil}
}

\maketitle
\IEEEpeerreviewmaketitle

\begin{abstract}
The optimization of viewers' quality of experience (QoE) in 360 videos faces two major roadblocks: inaccurate adaptive streaming and viewers missing the plot of a story. Alignment edit emerged as a promising mechanism to avoid both issues at once. Alignment edits act on the content, matching the users' viewport with a region of interest in the video content. As a consequence, viewers' attention is focused, reducing exploratory behavior and enabling the optimization of network resources; in addition, it allows for a precise selection of events to be shown to viewers, supporting viewers to follow the storyline. In this work, we investigate the effects of alignment edits on QoE by conducting two user studies. Specifically, we measured three QoE factors: presence, comfort, and overall QoE. We introduce a new alignment edit, named \textit{Fade-rotation}, based on a mechanism to reduce cybersickness in VR games. In the user studies, we tested four versions of fade-rotation and compared them with instant alignment. We observed from the results that gradual alignment achieves good levels of comfort for all contents and rotational speed tested, showing its validity. We observed a decrease in head motion after both alignment edits, with the gradual edit reaching a reduction in head speed of 8\% greater than that of instant alignment, confirming the usefulness of these edits for streaming video on-demand. Finally, parameters to implement \textit{Fade-rotation} are described.
\end{abstract}

\section{Introduction}
\label{sec:introduction}

The popularity of virtual reality (VR) is growing at an exponential rate. According to Cisco~\cite{Cisco18} and Huawei~\cite{Huawei17}, the traffic of this type of media is expected to increase 12 times between 2018 and 2023. Foremost, due to the availability of affordable Head-Mounted Displays (HMDs) and the attractiveness of realistic immersive experiences. As a consequence, the production of VR content (mainly in 360$^\circ$ video format) increases in quantity and quality. For example, the Cannes Film Festival~\cite{cannes} has a program dedicated to immersive technologies and cinematography. There are also several companies that specialized in VR film production, such as VR Gorilla, Planar, and Unreal Engine~\cite{vr-gorilla, planar, unreal-engine}.

To enable large-scale adoption of VR on-demand experiences, at least two questions arise. To what extent can content design improve users' quality of experience (QoE)? How can one optimize the delivery of resource-intensive 360 videos over regular internet transmission? Regarding the first question, cinematography studies present guidelines for QoE improvement~\cite{reilhac2016, milk2016chris, pillai2017grammar,argyriou2020design}, and viewers' engagement. Considering the second question, typical solutions apply adaptive bit rate (ABR) streaming. ABR schemes aim at allocating high-quality resolution only in the portion of the video where viewers gaze (a.k.a. viewport), optimizing the visual quality perceived by the end user, given the available network resources \cite{wang_twotier_2019, Chiariotti2021ASO}. Therefore, QoE optimization and streaming depends on viewport predictability and content relationship with user experience (UX). However, VR film production is in an exploration phase~\cite{reilhac2016, milk2016chris}. Understanding the impact of VR content on QoE is challenging, since QoE depends on numerous factors \cite{qoe_factors2022,Sharda2008CreatingAM}. For instance, systems and components factors (e.g., stalling, quality degradation), UX factors (e.g., feeling of immersion, discomfort), and content factors (e.g., content motion, genre, scene environment). The relationship between content spatio-temporal features and UX remains largely unexplored. Due to the high space of parameters to describe the content, the QoS, and the user perception together. Another challenging aspect is the high cost to conduct UX studies \cite{Borchert2020InVitro, Hassenzahl2006UserE, Sharda2008CreatingAM}.

From the cinematographic point of view,~\cite{brown2017coordinating} argued that it is critical developing a set of techniques that assist viewers in moving around the scene, following the intended storyline. Assuming that content is organized and/or manipulated to engage the audience, two types of viewing guidance techniques take place, the active and passive techniques \cite{Rothe2019GuidanceIC, Nassar2021EngagingBD}. The passive can use either a diegetic or non-diegetic attractor. Diegetic attractors are those elements included in the scene that makes part of the fictitious story, and it is inserted to capture the viewer's attention \cite{Nassar2021EngagingBD}. While non-diegetic attractors are elements outside the fictitious story, inserted in viewers' display, e.g., visual guidance effects like arrows, radar. On the other hand, active techniques support viewers in run-time, for example, by managing the camera to follow specific targets \cite{Hu2017Deep3P},\cite{Wang2020Transitioning360CN} or by manipulating the luminosity or saturation of the scene \cite{danieau2017attention}. Compared to active techniques, passive techniques are less restrictive; however, they do not suffice for full predictability required for streaming purposes.

Alignment edits actively rotate viewers' viewport to a specific region of interest (ROI) represented by a single point of interest (POI). To avoid intrusive rotations, short-duration rotations should be prioritized \cite{Wolf2021AugmentingTI}. Typically, alignment edits are triggered by a video player in run-time, so it is considered an active guiding technique. Alignment edits can be designed in various ways, combining rotation with visual effects or with motion transitions (translation, teleport). The impacts alignment edits have on users are not well understood. To our knowledge, \cite{dambra2018film} was the first to investigate alignment edits for cinematic VR (CVR) content, showing promising results when applied in adaptive streaming systems. However, alignment edits lack in-depth investigation, \cite{dambra2018film} presented analysis for a low number of contents, and considering one type of alignment edit, the instant alignment edit, where the rotation occurs in one frame interval. Finally, beyond CVR, alignment edits can be explored for VR applications~\cite{serrano2017movie}, and streaming application optimization\cite{sassatelli2018snap} reducing the bandwidth consumption, and network latency.

In this work, we investigate the effect of alignment edits on the viewing experience. Our goal is to gain insight into the usefulness and impact of these techniques on the viewer's QoE. Therefore, we conducted a series of experiments, according to each of the objectives of the study. First, we analyzed user opinions on their experiences, considering three experience factors (presence, discomfort, and QoE). Second, we investigated the user behavior for different alignment edit configurations, based on viewers' head motion.

From the exposed background, we claim that alignment edit is a useful technique, not fully investigated. Alignment edits amplify the accuracy of gaze prediction, detecting the situations in which it optimizes QoE, allows optimizing visual quality and story comprehension mutually, resulting in a QoE enhancement. Also, we claim that gradual alignment edits can be used in CVR without compromising the experience. To skip the expected compromising effects of gradual rotation \cite{dambra2018film}, we claim that combining gradual rotation with a cybersickness reduction technique is enough to make gradual alignment viable for CVR. Moreover, instant alignment edits amplify the odds of breaking the users' sense of presence \cite{Rahimi2020SceneTA}, similar to cuts in regular videos or teleport effect in VR \cite{Beck2021ApplyingDC,Mateer2017DirectingFC}. This points towards the necessity for a new alignment edit that can be more natural to the users, preventing the break in immersion. To cover those claims, this work focus on the following contributions:

\begin{enumerate}
    \item We provide collected subjective dataset from our user study, together with the analysis code to promote future research on the topic;
    \item We provide implementation description for a new gradual alignment edit technique based on based~\cite{Farmani2020EvaluatingDV}, to facilitate its incorporation in CVR applications;
    \item We include a new method for analyzing behavior based on alignment metrics that can be used in other similar studies;
    \item We provide recommendations for the implementation of instant and gradual edits to optimize QoE;
    \item We give insights for QoE optimizations based on the analysis of presence, comfort, and overall quality, crossing it with content analysis.
\end{enumerate}

We structure the paper as follows. In Section \ref{sec:relatedworks} we outline the background related to storytelling in 360-degree videos and present works about alignment edits. Section \ref{sec:methodology} describes the proposed gradual alignment edit and the experiment design. We describe the data preparation in \ref{sec:preanalysis}. Further, in Section \ref{sec:results} we present the results comparing both alignment edits, following the research hypothesis testing. Finally, Section \ref{sec:conclusion} includes the debate, recommendations, and limitations of our findings.

\section{Related Work}
\label{sec:relatedworks}

Following, we describe the cinematography aspects of immersive media, culminating in the alignment edits technique, as well as the challenges associated with implementing it as a QoE solution.

\subsection{Storytelling for 360 Videos}

Cinematography guidelines establish rules to achieve the feeling of continuity of the scene and, aesthetic goals. For example, to achieve scene coherency, a director should respect the ``180$^\circ$ rule,'' which restricts camera positioning across the action axis. Moreover, to achieve continuity of action, typically directors start action in one shot and immediately continue it after a cut. In addition, the 180$^\circ$ rule creates a virtual stage where the action unfolds~\cite{bordwell1997history}, and action cut simulates biological motion tracking processes~\cite{smith2008attentional}. 
Immersive media allow viewers to act as the camera, enhancing the freedom to explore the scene. Spatial displacements between ROIs in 360$^\circ$ videos introduce difficulties in creating a coherent narrative \cite{Weaving2021EvokeDS, young2000creating}, compared to regular videos. \textit{Continuity editing} holds in CVR \cite{serrano2017movie}, however many editing techniques (e.g., camera angles, zooms, fade, cut) may become ineffective in the 360$^\circ$ scenario, raising questions on how to create narratives for this type of immersive media. The development of new guidelines for immersive media support directors to improve user experience \cite{Fearghail2021UseOS}, \cite{Nassar2021EngagingBD}.

J. Brilhart \cite{brilhart2016} proposed an editing principle for CVR called Probabilistic Experiential Editing, a procedure that generates scene edits by estimating which areas of the content are more salient or perceptually important to the storyline. Another significant concept of CVR cinematography is the temporal and spatial density of the story \cite{Gdde2018CinematicNI} that corresponds to the quantity, positioning, and frequency of POI for a given story timeline. This study also examined viewers' tolerance for spatio-temporal story density. \cite{Aitamurto2021FromFT} examined variations in spatiotemporal viewing conditions in CVR, testing how it triggers the psychological condition of fear of missing out, resulting in anxiety and degrading viewer enjoyment. Cinematography studies argue that real-time editing could prevent plot missing and promote storytelling engagement in CVR \cite{brown2017coordinating,Nassar2021EngagingBD}. From a user study with 50 participants, authors from \cite{Gdde2018CinematicNI} found that for scenes with high spatial-temporal density, a significant part of the audience can miss the plot. For example, they found a case where 80\% of the participants could not correctly answer story-based questions such as ``What happened to the main character?'' ``Why did the character become aggressive?''

\subsection{Alignment edit strategies}
\begin{figure}[tb]
    \centering
    \includegraphics[width = 0.45\textwidth]{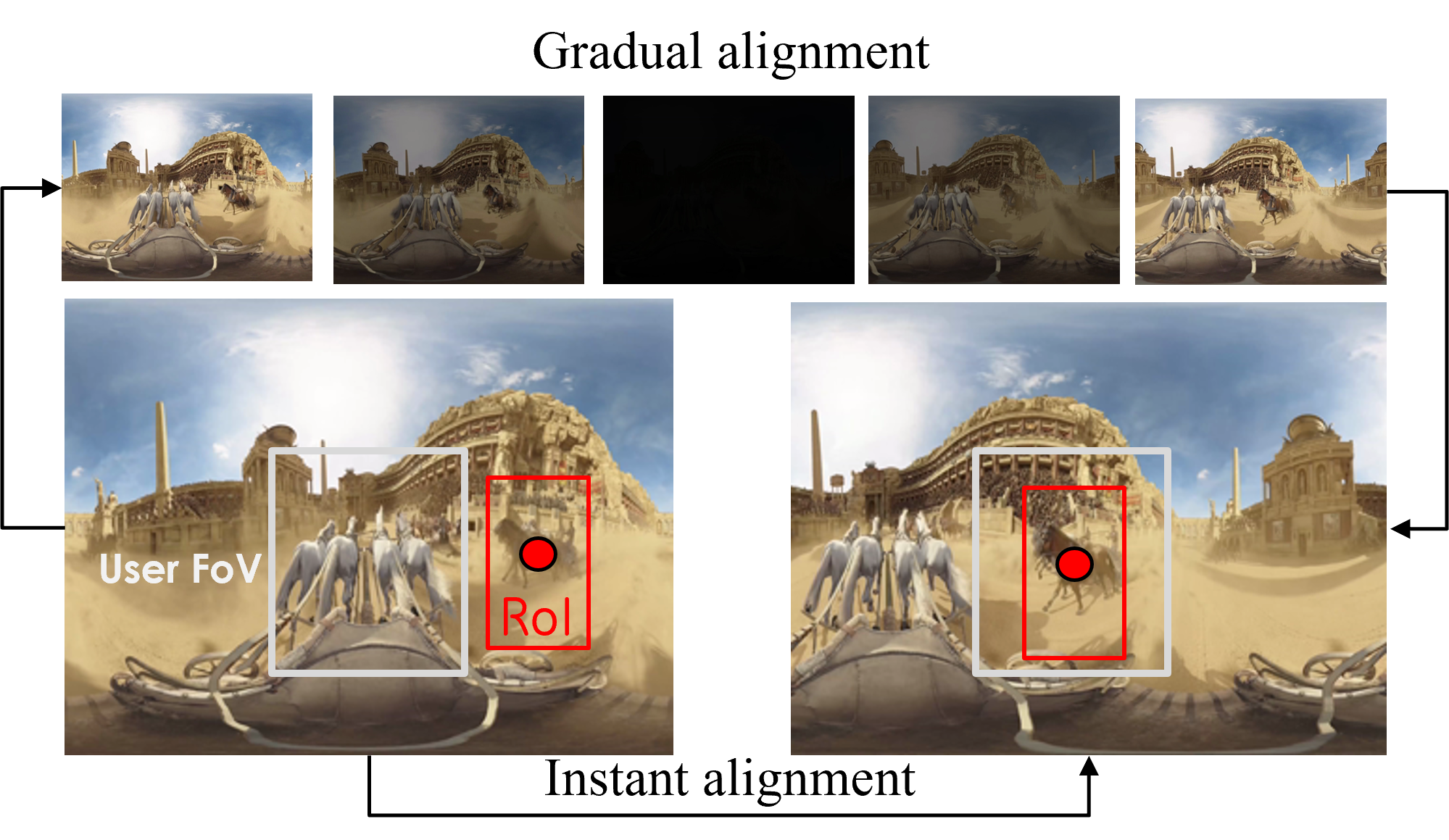}
    \caption{Gradual and instant alignment edits lining up viewer FoV with a specific ROI that otherwise would be missed. Gradual alignment combines content rotation with a rapid fade- in-out simulating blinking of the eyes, to make the edit as subtle as possible.    }
    \label{fig:Gradual}
\end{figure}

Recent user studies examined how video edits impact the viewing behavior for CVR content. 
In \cite{Cao2019APE}, the authors explored three types of transition effects (portal, fade, cut) and observed no conclusive reduction of story recall. Sitzmann \textit{et. al.}~\cite{serrano2017movie} studied head and eye-tracking data to examine the effects of content factors. They concluded that the number of ROI and, ROI displacement play a key role in user behavior. Specifically, the time to find ROI as well the stabilization of gaze in an ROI are related to those factors. Mara\~nes \textit{et. al.}~\cite{Maranes2020_VR_Cut_Behaviour} contributed in this direction by researching the impact of the number of ROIs before and after cuts, proposing important area-based behavior metrics for head tracking data. The work of Kjaer \textit{et. al.}~\cite{kjaer2017can} explored the effects of editing by adhering to both the principles of attention and of match-in-action.  Speicher \textit{et. al.}\cite{speicher2019exploring} suggested visual guidance techniques as a post-production resource that can be implemented in video players to expand accessibility~\cite{montagud2020culture}. Another approach, in the montage or post-processing of the content, is to manipulate/edit footage aligning the potential POI across shots. In the work of Pavel \textit{et. al.}~\cite{pavel2017shot}, the authors analyze an additional shot orientation technique that helps viewers visualize all the critical information in 360-video stories. Finally, Sassatelli \textit{et al.}~\cite{sassatelli2018snap} proposed a technique, aimed at snapping instantly viewers to ROI, this alignment technique allows the VR content creator to drive the user's attention. Figure \ref{fig:Gradual} illustrates how gradual alignment works as an alternative to instant alignment.

These techniques manipulate the viewer’s gaze orientation by rotating the camera or the visual sphere using pre-defined parameters. Studies in this field closely relate user studies with content manipulation, for example, \cite{kjaer2017can} studied the effect of cut frequency on viewer’s disorientation and found evidence that editing does not pose a problem in relation to cinematic VR. Real-time edits introduce opportunities for customized optimizations. For instance, adaptively supporting accommodation of people susceptible to cybersickness \cite{Garrido2022FocusingOC,Tian2022ARO}, or prone to diverge from a pre-designed storyline (individuals with low reaction time \cite{Eom2019DevelopmentOV}). Furthermore, they also enable system optimizations, e.g., expanding the efficacy of streaming applications. The authors of \cite{dambra2018film} and \cite{sassatelli2020new} examined how the instantaneous alignment technique reduces exploratory behavior and improves streaming metrics. Furthermore, an important insight from \cite{dambra2018film} is that instant edits can be imperceptible to viewers, showing that, if inserted in specific content conditions, edits can even not be noticed while incorporating technical improvements. Moreover, real-time edits have a set of applications aiming to reduce cybersickness, especially for VR games context \cite{Shi2021VirtualRS,Lin2020HowTP,Yildirim2019DontMM,Teixeira2020EffectsOD}. Closer to our study interests, the authors of \cite{Farmani2020EvaluatingDV} proposed a technique that acts on gaze direction. To avoid activating cybersickness, when a rapid head movement happens, the system triggers this technique. When triggered, users' screen illumination decreases while discrete angular offsets (of 25$^\circ$) occur. This bioinspired solution simulates blinking of the eyes, and a UX experiment proved the reduction of cybersickness by up to 40\% for a first-person VR shooting game.

\subsection{Alignment edit challenges}

The 360-degree video viewing task is a psychophysiological phenomenon, underlying the human visual system. In this task, viewers have two main behaviors: exploratory, where the gaze navigates the content freely searching for some interesting spot, and fixation, the act of focusing on some region of interest (ROI). This selective process is underlying the sequence of gaze directions \cite{Rossi2017NavigationawareAS,Maranes2022}. Deciding how to align ROIs in a scene is a semantic choice; in consequence, if something captures user attention in the wrong way that filmmakers expect, viewers can miss notable events that help understand the story and enjoy the best from the content, thereby likely degrading the user experience. In a nutshell, the higher the degree of freedom of 360$^\circ$ videos, the higher the odds to enable the sensation of immersion, albeit it also amplify the probability of missing notable events \cite{Aitamurto2021FromFT}.

Relative motion between content and user gaze is a research interest in VR, it can be the most significant factor in users’ QoE \cite{Jerald2008SensitivityTS}. The sensation of self-movement is crucial to understanding relative motion effects. It is known that the interaction between visual and vestibular systems triggers self-motion, for example, in flying scenes or when a virtual object crosses the viewport \cite{Litleskare2019CameraSI}.  \cite{Tian2022ARO} extensively investigated visually induced motion sickness.  \cite{Serrano2020ImperceptibleMO} measured just noticeable differences in the head lateral shifts, showing that it is possible to implement real-time optimizations based on content parameters, specifically object distances, that would lead to an imperceptible translation gain.  

In particular, the activation of cybersickness through motion scenes imposes one of the main challenges to QoE optimization, since it can degrade the sense of presence and lead to discomfort. Therefore, it is essential to consider the trade-off between cybersickness and presence when designing any mechanism that acts in viewers’ sense of motion \cite{Dorado2015MethodsTR}, \cite{Weech2019PresenceAC}. Matching in real-time both content and viewer motion is challenging, although critical since it can encourage spatial presence and empower the psychological experience of spatial fusion \cite{Tian2020SpatiotemporalEM}. 

The authors of \cite{sassatelli2018snap} argue that the potentiality of cybersickness activation in CVR contexts should limit mechanisms based on gradual rotation. However, to the best of our knowledge, no empirical test satisfy this assumption directly. Besides the potentiality of comfort impairment, the gradual rotation has advantages in terms of immersion; by embedding scene transition with scene motion, it is expected to conserve the sense of presence. Moreover, an important assumption of this work follows evidence from similar studies \cite{Farmani2020EvaluatingDV}, \cite{Eftekharifar2021TheRO},\cite{Ang2020GingerVRAO} that gradual rotations can be tuned to be as comfortable as instant alignments. Although often CVR has multiple ROIs in a scene, in this study we considered the simplest case of one ROI alignment with the viewport, the problem of \textit{i.e.,} we do not consider the multiple ROI alignment problem.

\section{Experiment Design Setup}
\label{sec:methodology}

We aim to measure alignment edit impacts in QoE and behavior.
We propose an editing technique, named \textit{fade-rotation}, that serves as a gradual alignment, designed based on a VR gaming technique from previous work \cite{Farmani2020EvaluatingDV}. The instant alignment edit included in the study, called \textit{snap-cut}, was first implemented by \cite{dambra2018film}. The study compared two versions of snap-cut: offline (inserted in the video previous to its playback) and online (triggered in video run-time). However, they did not cover the comparison between snap-cut and other alignment edit strategies.

The optimization of QoE in CVR depends on resolving the trade-off between the sensation of presence and comfort  \cite{Weech2019PresenceAC}. It is crucial to understand in which conditions the alignment edit selection 1) maximizes the desirable user sensation of “presence”, and 2) minimizes the undesirable feeling of “discomfort”. Moreover, we have to confirm if our new gradual alignment edit provides the main advantages of instant edit, reducing head motion after the edit, and preventing users from missing important events. Therefore, we formulate the following hypothesis to conduct this study:

\begin{enumerate}
    \item [$\mathbf{{ H1:}}$] The degree of comfort of gradual edit (fade-rotation) is equivalent to that of instant edit (snap-cut);
    \item [$\mathbf{{ H2:}}$] The instant edit (snap-cut) has a higher negative effect in presence than gradual edit (fade-rotation);
    \item [$\mathbf{{ H3:}}$] The QoE factors of presence and comfort influence the alignment performance of the edit;
    \item [$\mathbf{{ H4:}}$] The alignment edits reduce viewers’ head speed after the edit.
\end{enumerate}

\subsection{Experiment procedures}
\label{sec:procedures}
\begin{figure}[h]
    \centering
    \includegraphics[width = \linewidth]{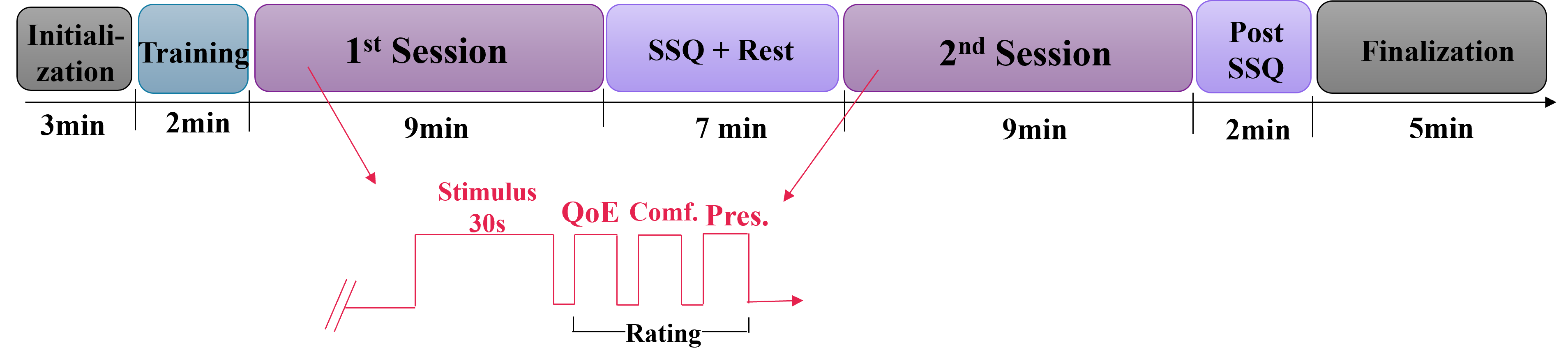}
    \caption{Procedure of the experiment, and the subject rating time structure.}
    \label{fig:procedure}
\end{figure}

We conducted a within-subject user study in which participants evaluated each video in terms of QoE, discomfort, and presence as shown in Figure \ref{fig:procedure}.
The experiment methodology was equivalent to the Absolute Category Rating with Hidden Reference (ACR-HR) of quality studies according to ITU-T recommendation P.919~\cite{itut2021VRexp}. Users assessed the manipulated and reference videos (without editing) for all contents as a single stimulus. Subjective responses conformed into a 5-point scale single-item questionnaire for QoE, discomfort \cite{Prez2018Comfort}, and presence \cite{Bouchard2004Presence}. The questionnaire design was useful to avoid misinterpretations, fitting the number of stimuli in the 30-minute tolerance limit~\cite{itut2021VRexp}, and facilitating the execution of reliability checks in the data \cite{Borchert2020InVitro}. Table \ref{tab:scales} present questions and scales answered by participants.

\begin{table}[tb]
\small
\centering
    \caption{Subjective assessment measures.}

    \begin{tabular}{c  c  c }
    \toprule
    \textbf{Questions} & \textbf{Scale} & \specialcell{ \textbf{Measure} \\ \textbf{Reference}}  \\
    \midrule
    
    \specialcell{Are you feeling any sickness \\ or discomfort now? \\ In the below \\ check the appropriate value.} & \specialcell{(1) Unbearable \\ (2) Unpleasant \\ (3) Uncomfortable \\ (4) Light effects \\(5) No Problem} & \specialcell{Comfort \\ \cite{Prez2018Comfort}} \\
    \midrule
    \specialcell{To which extend do you feel\\  present in the virtual \\ environment, as if you \\ were really there?} & \specialcell{(1) Nothing \\ (2) Little much \\ (3) Reasonably \\ (4) Very much \\ (5) Entirely} & \specialcell{Presence \\ \cite{Bouchard2004Presence}}  \\
    \midrule
    \specialcell{Evaluate the overall \\quality of the video \\ In the below \\ check the appropriate value.} &
    \specialcell{(1) Bad \\ (2) Poor \\ (3) Fair \\ (4) Good \\ (5) Excellent} & \specialcell{Overall \\ Quality \\ \cite{itut2021VRexp}} \\
    \midrule
    \specialcell{Evaluate the following \\symptoms: \\ Nausea, Vertigo, Sweating, \\ Stomach  awareness, \\ Increase in salivation, \\ Difficulty in concentration.} &
    \specialcell{For each\\symptom: \\(1) None \\ (2) Slight \\ (3) Moderate \\ (4) Severe} & \specialcell{Cyber-\\sickness \\ \cite{Sevin2019PsychometricEO}} \\
    \bottomrule
    \end{tabular}
    \label{tab:scales}
\end{table}

The experiment contained six videos and six edit types, corresponding to a total of 36 conditions. We include two types of edits: instant alignment edit (fade-rotation with four different angular speeds 10$^\circ$/s, 20$^\circ$/s, 40$^\circ$/s, and 60 $^\circ$/s), instant alignment edit (snap-cut), and a baseline without edit. Due to a technical issue with the first device, we carried out the experiment on two occasions. The first sub-experiment (with Oculus Rift S) took place in August 2021 and the second (with Meta Quest 2) in November 2021.

We conducted the experimental procedures following ITU-T P.919, the most recent recommendation for subjective experiments with 360 videos \cite{itut2021VRexp}. Table \ref{tab:population} presents the pool of the participants for each sub-experiment. We recruited 40 and 23 participants for the first and second experiments, respectively. The sampled population had a wide variety of ages and HMD experience, and the proportion of women was greater than 40\% in both experiments. In total, we collected 6804 opinion scores, and  1300-2000 head tracking samples per video watched.\footnote{Given publication acceptance, our dataset will be made available for the scientific community.}
Participants were invited on social media~\footnote{https://lucas-althoff.github.io/Experimento-FadeRotationVR/}~\footnote{In correspondence with the ethical board, Covid-19 safety protocols were applied to gather participants within the university}. We prioritize recruiting participants from outside the university to improve the population sampling.
A full run of the experiment lasts approximately 37-40 minutes. During the test, participants were seated on a swivel chair. Participants who wore glasses or lenses kept them on throughout the session. As shown in Figure \ref{fig:procedure}, the experiment had four phases. The experimental session was completely automated, and no intervention from the experimenter was required, except for some clarification in the instruction phase if participants needed it. The pre-questionnaire was based on VQEG's~\footnote{https://www.its.bldrdoc.gov/vqeg/projects/immersive-media-group.aspx}. At the end of each trial, participants gave their scores on the score interface using the HMD controller. After the first session, the participants completed a cybersickness questionnaire, remove the HMD, and had a five-minute break to prevent bias and cognitive load \cite{itut2021VRexp}.

In the Rift experiment, participants completed the SSQ and the pre-/post-questionnaires using a computer and keyboard. While in the Quest 2 experiment, participants filled in information using the HMD controller. For each participant, the stimulus order was randomized in two sets, no fast fade-rotations were included for the first 8 watched videos (angular speed equal to 40$^\circ$/sec and 60$^\circ$/sec). This condition prevents bias from premature activation of cybersickness~\cite{Farmani2020EvaluatingDV}.

\begin{table}[h]
\centering
\caption{Experiment population summary for both devices.}
    \begin{tabular}{l  c  c  c  c  c  c  c }
    \toprule
    \textbf{Device} & \textbf{Num.} & \multicolumn{4}{c}{\textbf{Age}}& \textbf{Prop.} & \textbf{1st time}\\
    &\textbf{part.} & \textbf{Avg.}&\textbf{Std.}& \textbf{Min.}&\textbf{Max.} & \textbf{women} & \textbf{w/ VR} \\
    \midrule
    Rift S &40 & 35.62 & 14 & 15& 65& 60\% & 55\% \\
    \hline
    Quest 2 &23 & 29.56 &  7& 18& 41& 43\% & 60\% \\
    \hline
    \textbf{Total} & 63 & 33.4 &  11& 15& 65& 53.8\% & 56.8\% \\
    \hline
    \end{tabular}
    \label{tab:population}
\end{table}

\subsection{Video stimuli}
\label{sec:stimuli-parameter}
To test edits under a variety of conditions, we classified video stimuli into three categories of camera motion types (static, steady, and dynamic). We consider a video as "steady" when no prominent camera acceleration occurs independently of direction \textit{for the majority of the scene}. The "dynamic" class is the opposite. This classification corresponds to our previous knowledge that camera acceleration introduces dynamic content motion, a relevant factor that prompts cybersickness \cite{Litleskare2019CameraSI, Eftekharifar2021TheRO}, and has a significant impact on QoE.

To prepare video stimuli from video sources, we extracted the soundtrack, and the transitions had the same duration and started at the same time-stamp to prevent bias related to content structure. Videos were encoded with the H264 codec, 3840x1920 resolution, EPR projection, 40 kbps target quality and 60 fps. For the aligned versions, the same angular rotation was applied. Edits were applied in the source videos manually using Adobe Premiere software and VR projection plugin, specific for CVR editing.

For the selection of stimuli, no criteria were used for the ROI number, and we included both outdoor and indoor content to evaluate the effects on a wide range of content. The audio of the videos was removed to avoid generating attraction bias. In the preparation of the stimuli, we looked for scenes where the reference region, the central point, had important and actively moving objects, so that viewer attention was retained until the intervention occurred.

\begin{figure*}
    \centering
    \includegraphics[width=.8\paperwidth]{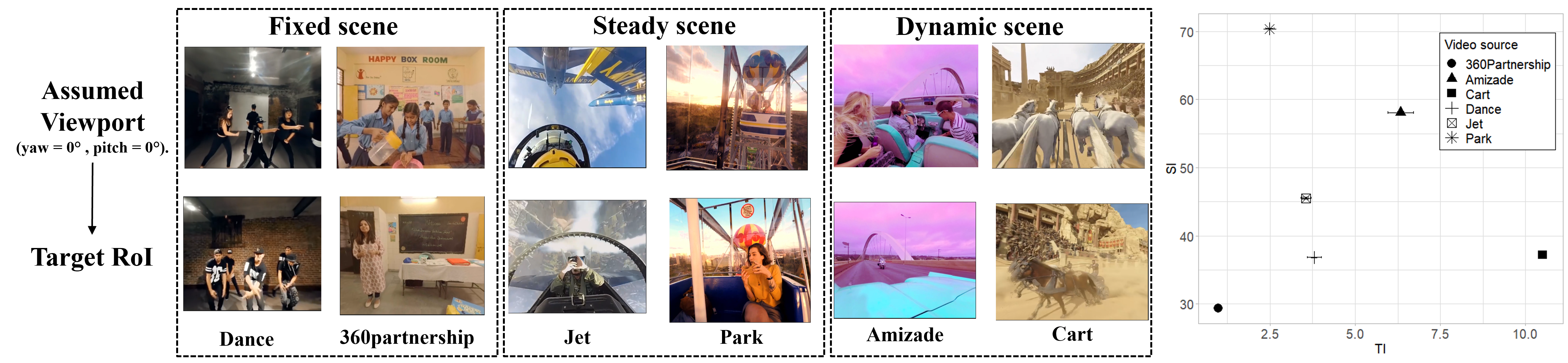}
    \caption{Video-stimuli for subjective experiment. Left: Video source frames. Right: Spatial and temporal indexes of videos}
    \label{fig:siti}
\end{figure*}

The set of video stimuli covers a wide range of spatial and temporal information. Spatial and temporal structural indices of each video are shown in Figure \ref{fig:siti}. Of the six videos selected, four were gathered from public datasets Directors Cut \cite{knorr2018director} ``360partnership,'' UTD \cite{Nasrabadi2019ATA} ``Jet,'' ``Dance'' and ``Cart,'' and the other two ``Amizade'' and ``Park,'' were provided to our research by the filmmakers.

\subsection{Alignment edit design}
The proposed \textit{fade-rotation} edit is distinct from the original \cite{Farmani2020EvaluatingDV} since it is activated on a pre-selected video time-stamp, while in the original version, the trigger was controlled by head speed \cite{Farmani2020EvaluatingDV}. Selecting a prior time-stamp to perform an edit is a crucial feature for automating edits. In fade-rotation, the rotation is combined with a fade-out-in effect that delivers a smooth transition to attenuate or avoid cybersickness. This edit simulates the natural blinking of the eyes. On the other hand, \textit{snap-cut} is implemented between two frames of the video acting as a cut, but also rotating the visual sphere in the horizontal direction.

\begin{figure}[tb]
    \centering
    \includegraphics[width = 1.\linewidth]{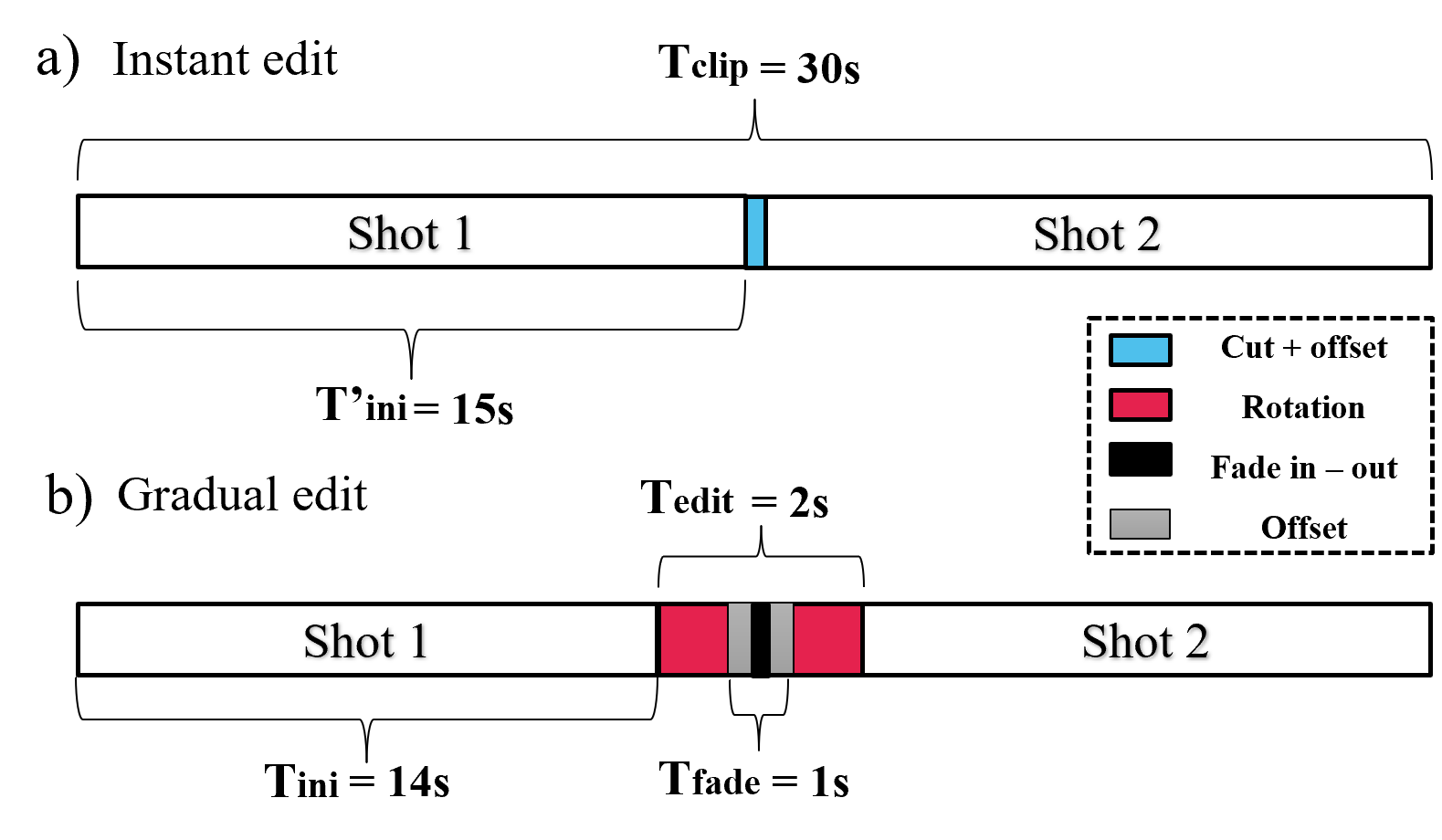}
    \caption{Edit parameters for the alignment edit a) fade-rotation b) snap-cut. Parameters settings applied in the video-stimuli is presented.}
    \label{fig:FR}
\end{figure}

The temporal structure of both edits is shown in figure \ref{fig:FR}. To avoid bias, all stimuli have the same length ($T_{clip}=30s$), and the same interval of visible information before changing shot (15s). So, if two participants were observing the same location when the edit start, then they would end up observing the same target regardless of the type of edit applied. Fade-rotation has five parameters, the video time-stamp where it is triggered ($T_{ini}$), the edit duration ($T_{edit}$), linear fade-out-in duration ($T_{fade}$), the whole angular displacement ($\theta_{r}$), and the offset angle in the black frame ($\theta_{off}$). For this study, we fixed $(T_{ini} = 14s ; T_{edit} = 2s ; T_{fade}=200ms)$ and manipulated $\theta_{edit} ; \theta_{off}$ to test fade-rotations with different angle speeds $\omega = \{10, 20, 40, 60^{\circ /s}\}$. The introduction of this spin offset is mandatory to control the angular speed while keeping $T_{edit} = T'$ fixed. The amount of angular offset $\theta_{o}$ depends on:  
\begin{equation}
    \theta_0(\theta_r, \omega) = \omega T' - \theta_r(video),
\end{equation}

where $\theta_r(video)$ are the alignment displacements for a specific video. This displacement is measured as the angular distance between the center of the visual sphere and the ROI at $T = T_{ini}$.

The snap-cut has only two parameters, angular rotation ($\theta_{sc}$) and trigger time-stamp ($T'_{ini} = 15s$). The angular displacements for both edits are the same ($\theta_{sc} = \theta_{r}(video)$) in a period of one frame.

Offline alignment edits depend on the assumption of the gaze position before the edit. The assumed pre-edit viewport is centered at position yaw=0$^\circ$, pitch=0$^\circ$. This point is known as the center point, the same point where the user starts to watch the video, and the initial stable head posture. In general, the gaze of VR viewers is biased to the central point, independently of the content \cite{serrano2017movie}. We consider two facts to improve the chance that the viewers stabilize at the desired center point at $T_{ini}$. First, all videos had an ROI at the center point at $T_{ini}$. To that, we had to manipulate the initial center point of two videos (Park = -180 $^\circ$, Dance = -86 $^\circ$). Second, the length of the shot was defined to cover the average accommodation time of 14-16s, reported by \cite{serrano2017movie}. The snapping of the sphere of both edits was applied exclusively in the yaw direction, as recommended in \cite{dambra2018film}. The displacement angle $\theta_r(video)$ of each video was: Park = 180$^\circ$, Jet = 180$^\circ$, 360Partnership = 170$^\circ$, Dance = 86$^\circ$, Amizade = -120$^\circ$, Cart = 120$^\circ$.

\subsection{Materials}
\label{sec:platform}
All data were collected with the procedure described in \ref{sec:procedures}, and it was implemented using a platform specially developed to guarantee uniformity of the data and flexibility of our experiment design. Since all solutions available for subjective evaluation of 360$^\circ$ videos require closed source game engines, we developed this platform to perform the experimental procedures. In the following, we briefly describe the monoscopic 360$^\circ$ video subjective assessment tool (Mono360) and the experiment equipment.

Mono360 was built using a client-server architecture based on the HTML 5 standard. The back-end of the application executes the \textit{PHP} based \textit{Yii framework} on the server side. The database are \textit{Postgresql}, and the front-end interface is based on several \textit{JavaScript} frameworks and libraries. With this web-based solution, it was possible to input data through the HMD browsers that support the \textit{WebXR} API.

We used the Three.js\footnote{https://threejs.org/} library to render the 360$^\circ$ video, wrapping a \textit{WEBGL2} renderer. The rendering procedure consists of decoding the video texture into two spheres for both eye screens. The decoding process is managed by the browser. The interface elements were built using the \textit{Three-mesh-ui} library. Since all rendering procedures were customizable, we were able to adjust both sub-experiments to have the same behavior across different devices. Data were stored as screen-normalized coordinates using equirectangular projection. To collect head tracking data, we define the gaze direction as the following: the normalized screen position (x,y) of the intersection between the center of the viewport and the video texture. We used the origin convention as the top left corner. Mono360 stores user data after the end of each video rating, so it can be resumed at any time if an adverse condition occurs. All communication with the server is asynchronous. The videos from the session are pre-fetched into the browser before playing, this feature guarantees that videos were played without stall.

\section{Data Preparation}
\label{sec:preanalysis}

The analysis considers alignment edits as distortion in the reference Source Video Sequence (SRC), as typical in quality assessment studies\cite{itut.p910.1999}. The parameters to build the Processed Video Sequences (PVS) are described in Section \ref{sec:stimuli-parameter}. We organize the analysis into two parts. First, analysis of user QoE using subjective rating data. Second, is the analysis of user behavior using head tracking data. Finally, the pooling of the two sub-experiments is described before analyzing the data.

\subsection{Subjective data}
The distribution of subjective scores can be represented by a vector
$\mathbf{x^j} = (x_1,...,x_5)$ of scoring elements $x_i \geq 0$ with 5-point scale $i = \{1,2,...,5\}$, where $j = \{P,C,Q\}$ maps the response to comfort ($j=C$), presence ($P=2$) and overall QoE ($Q=3$) respectively. From distributions, we compute the mean opinion score (MOS) evaluations for each response $j$ depending on the pool of $N$ participants:

\begin{equation}\label{eq:meanScore}
\text{MOS}^j = \frac{1}{N} \sum_{i=1}^{5} x^j_i. \\
\end{equation}

On the other hand, the standard deviation of opinion score (SOS) for each factor is given by

\begin{equation}
SOS^j = \sqrt{\frac{x^j_i\cdot (i - MOS^j)^2}{(\sum _{i=1}^5 x^j_i) - 1}}.
\end{equation}

For simplicity, we consider for now on a margin of error equal to 0.05, and a 95\% confidence for the statistical tests. In turn, the confidence interval for the distributions is

\begin{equation}\label{eq:ciScore}
\text{CI}^j_{95\%} = \text{MOS}^j \pm \dfrac{1,96 \cdot \text{SOS}^j}{\sqrt{N}}.
\end{equation}

\subsection{Pooling Sub-experiments data}

First, we performed a Welch's t-test to identify if the presence scores ($MOS^P$) of the sub-experiments are statistically different. For that, we perform a pairwise comparison between two sub-experiment groups (Rift and Quest 2 devices). We chose this test because the sample sizes are different. Both t-tests show no statistically significant differences between the $MOS^P$, with $p >> 0.05$, therefore there is no difference in presence by using Rift or Quest 2 devices. We performed the same procedure for the comfort scores ($MOS^C$). Initially, we identified a statistically significant interaction considering the whole group. However, when performing pairwise comparison splitting data by edit type, we observed a significant difference only for one group. The group with gradual edit with 20$^{\circ}$/s. However, for all other aggregations, no significant differences were observed. From these results, we pool together sub-experiment datasets with the reminder that comfort scores which include 20$^{\circ}$ have device-related bias.

\subsection{Head motion data}
\label{sec:HM_metrics}
To quantify edit alignment accuracy, we must measure the alignment between participants' viewport and target ROI, for that we need a distance measure. \textit{Orthodromic Distance} is the most suitable measure to compute distance for spherical surfaces, when compared to the classical \textit{Mean Squared Error} and \textit{Angular error} \cite{RomeroRondon2021TRACKAN}. This is because it can handle the periodicity of the latitude while fitting better than other distance metrics to the spherical geometry distance problem. Furthermore, \cite{Rossi2019SphericalCO} showed that orthodromic distance is a reliable proxy of viewport overlap.

\subsubsection{Pre-processing HM data}
To prepare raw data, we first convert it from normalized screen coordinates ($X$,$Y$) into Eulerian coordinates ($\phi$,$\theta$), then translate the origin into screen center [0.5,0.5], and then we rescale the coordinates from [0,1] to $\phi \in [0, \pi]$ and $\theta \in [0,2\pi]$. From Euler angles ($\phi$, $\theta$) we compute the 3D Cartesian points of the spherical surface. After this coordinate transformation, each data point has the form $u = (x,y,z,t)$, where t is the video time-stamp of the collected values. With the transformed data, we can compute the orthodromic distance:

\begin{align}
    \label{eq:circ_dist}
    d(u,u') = 2R \cdot arcsin(\dfrac{c(u,u')}{2R}),
\end{align} 

where $c(u,u') = \sqrt{(x-x')^2+(y-y')^2+(z-z')^2}$ is the Cartesian distance between two points on the spherical surface $u, u'$. Despite great advances in head tracking data reliability, devices are still vulnerable to a series of problems, \textit{e.g.}, drift, tilt, and stationary jitter \cite{LaValle2014HeadTF}. So, it is important to apply data correction to skip sampling error or inaccuracy in the alignment measures, then we implemented mean sampling.

The frequency of HM data fluctuates over time. Therefore, our mean sampling must depend on the amount of data collected in each time interval considered. After pre-processing, each data point is a vector with 3D positions and a temporal index $u = (x,y,z,t)$.

\subsubsection{Alignment metric}

For the $i$th video watched by the $j$th participant, assume the pairwise \textit{mean distance} in a given time-stamp $t_0$ as the mean of all single distances within the interval $[t_0,t_0+\Delta t]$, where $\Delta t$ is a fixed sampling time interval. What leads to

\begin{equation}
    \label{eq:align}
    D_{ij}(u,v,t_{0}) = \dfrac{1}{N_{ij}(t_{0})} \sum_{k=near(t_{0})}^{near(\Delta t + t_{0})}d(u_{k},v_{k}),
\end{equation}

Where $d$ are given in equation \ref{eq:circ_dist}, $N_{ij}(t_{0})$ is the number of samples in the considered time interval, and the discrete time index $k$ computed as $near(t') = argmin(|k \Delta t_{ij} - t'|)$ for a given continuous time $t'$.

Our analysis will depend on the computation of the alignment metric. Align calculated from equation \ref{eq:align}, with the target ROI $ROI_{ij} = (x,y,z,t)$ and the actual viewport position $FOV_{ij}(x',y',z',t)$. From $D_{ij}$, the alignment edit performance for each trial is given by:

\begin{equation}
    A(FOV,ROI)_{j} = D_{ij}(FOV,ROI,t) < \tau,
\end{equation}

where $\tau$ is the tolerance of the angular distance to be considered ``aligned." For all videos used in the experiment,  the target ROI was put at the center point at the end of the edit, so $ROI_{ij} = (0,0)$, so the only free parameter to compute $A(\cdot)$ is the user position $FOV_{ij}(x',y',z',t)$.

We fixate the mean sampling interval as $\Delta t=250$ms (equivalent to approximately 30 pairwise samples for typical data sample frequency), $t_{0}=15$ s for snap-cut, and $t_{0}=16$ s for fade-rotations. Since the viewport adopted by our study is 90$^\circ$, we considered $\tau = 60^\circ$. In addition, the angular $A = 60^\circ$ mean approximately 30$^\circ$ of viewport overlaying \cite{Rossi2019SphericalCO}, and both devices used has 90$^\circ$ FoV \cite{Younis2019AHD} \cite{Kelly2022DistancePI}. 

\begin{figure}
    \centering
    \includegraphics[width=0.55\linewidth]{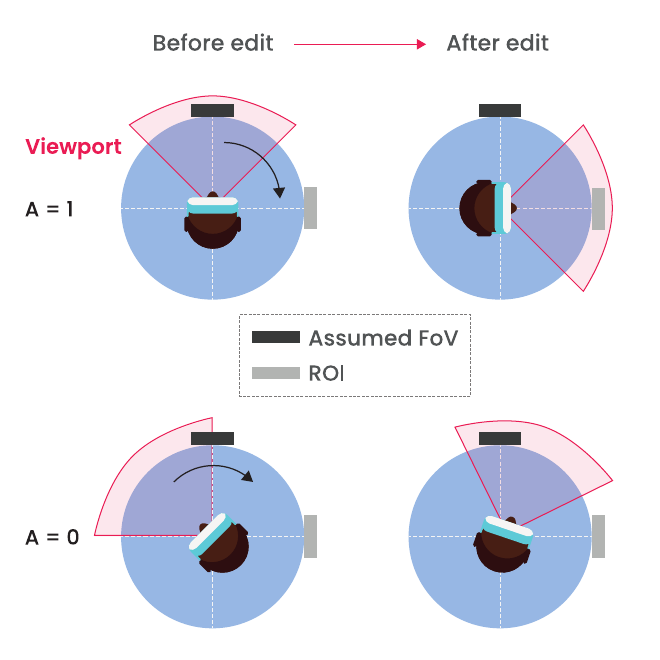}
    \caption{Possible states of alignment, $A = 1$ (first row) when alignment is successful, and $A = 0$ (second row) otherwise. The mean distance between viewport and ROI 1s after edit is used to compute the $A$. We applied a distance threshold of $\tau < 60{^\circ}$ to classify each trial in terms of $A$.}
    \label{fig:align}
\end{figure}

Binary classification of alignment between the viewport and ROI $A \in [0,1]$ (aligned, nonaligned) was considered, and is illustrated in Figure \ref{fig:align}. Makes it possible to analyze the state of alignment across the entire video. We are specifically interested in the short term before and after the edit. For example, we detach those participants that were already aligned but were misaligned by the edit with those that happened the opposite way. For the analysis of the state of alignment, we classified each trial of each participant into a dummy variable $\{1,2,3,4\}$ representing the four combinations between states (aligned, non-aligned) before and after the edit.

\subsubsection{Head speed metric}
\label{sec:hm}
Similarly, head speed $V_{ij}$ is computed from HM data by averaging a list of head speed samples over a specific time interval represented by $N$ samples.

\begin{align}
\label{eq:speed}
    V_{ij}(t_{0}) = \dfrac{1}{N_{ij}(t_{0})} \sum_{k=near(t_{0})}^{near(\Delta t + t_{0})}\dfrac{d(u_{k},u_{k+1})}{w(k+1)-w(k)},
\end{align}

where $w$ is the list of time-stamps indexed by $k$, and $d$ is calculated by the orthodromic distance of eq. \ref{eq:circ_dist}.

\begin{figure}
    \centering
    \includegraphics[width=0.9\linewidth]{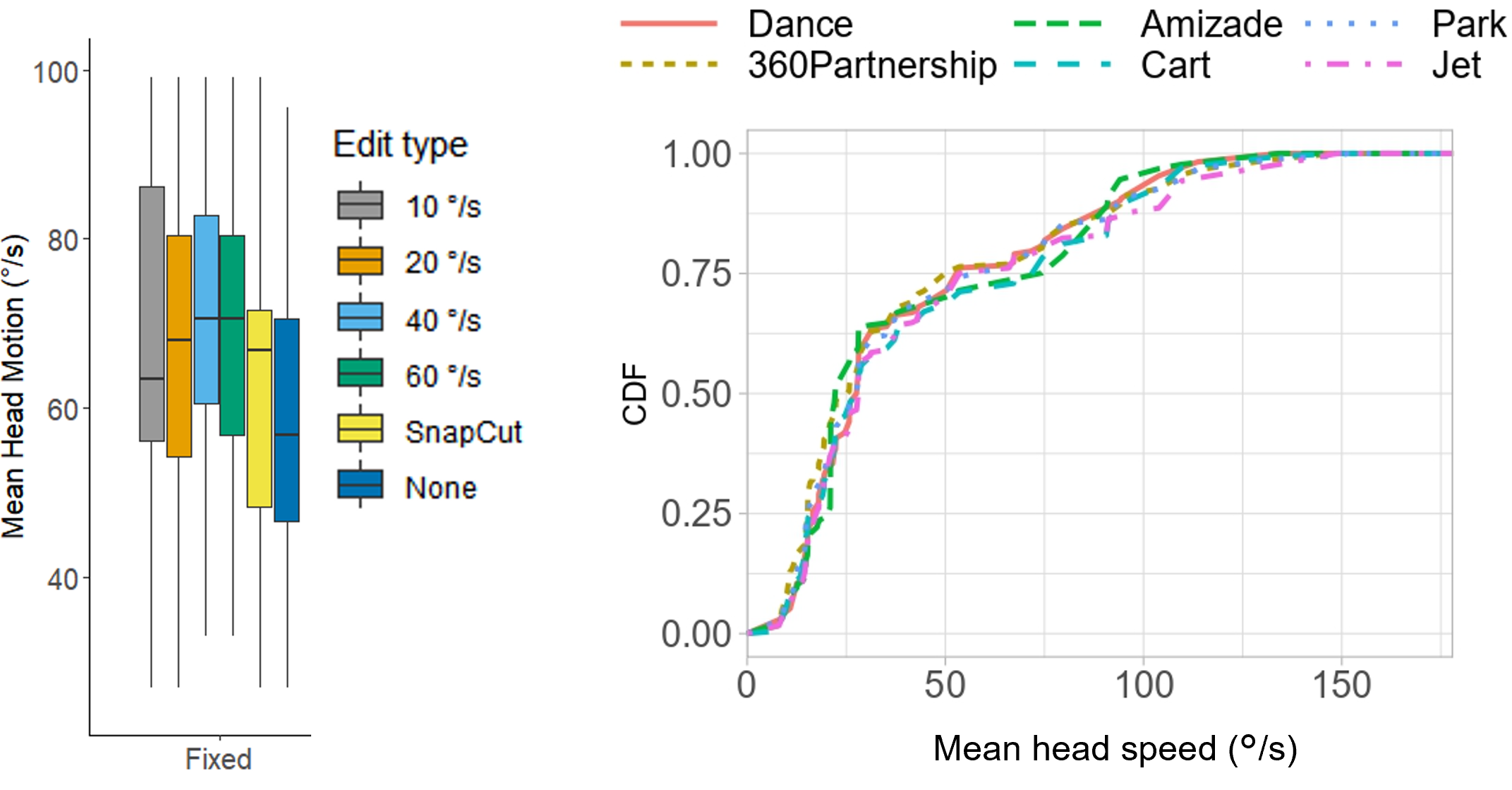}
    \caption{Distribution of head speed across data. Left: Boxplot for different edit type for fixed videos. Right: CDF of the mean head speed for all contents.}
    \label{fig:hmspeed}
\end{figure}

After mean sampling head speed, we must remove outliers with exotic high head speed to perform the analysis. This could have happened due to wrong measures. From the CDF of head speeds across all contents, shown in Figure \ref{fig:hmspeed}, we defined the threshold of 150 $^\circ$/s, as this velocity explains most of the data. Values greater than 150 $^\circ$/s are filtered.

\section{Results}
\label{sec:results}

In the following, from data collected in both sub-experiments, we analyze the effects of gradual and instant alignment edits on users' behavior and perceived QoE. To address all the research hypotheses presented in Section \ref{sec:methodology}, two targets comprise our analysis, the Source Video Sequence (SRC) and the Processed Video Sequence (PVS), depending on the hypothesis analyzed.

\subsection{Evaluation of QoE subjective ratings}

Beginning with the exploratory analysis, we will examine 1) the score distributions in terms of comfort, presence and overall QoE 2) the analysis of the MOS by video 3) the observed data patterns 4) the correlation between scores 5) the analysis of the MOS by camera motion class.

\begin{figure}
    \centering
    \includegraphics[width = .9\linewidth]{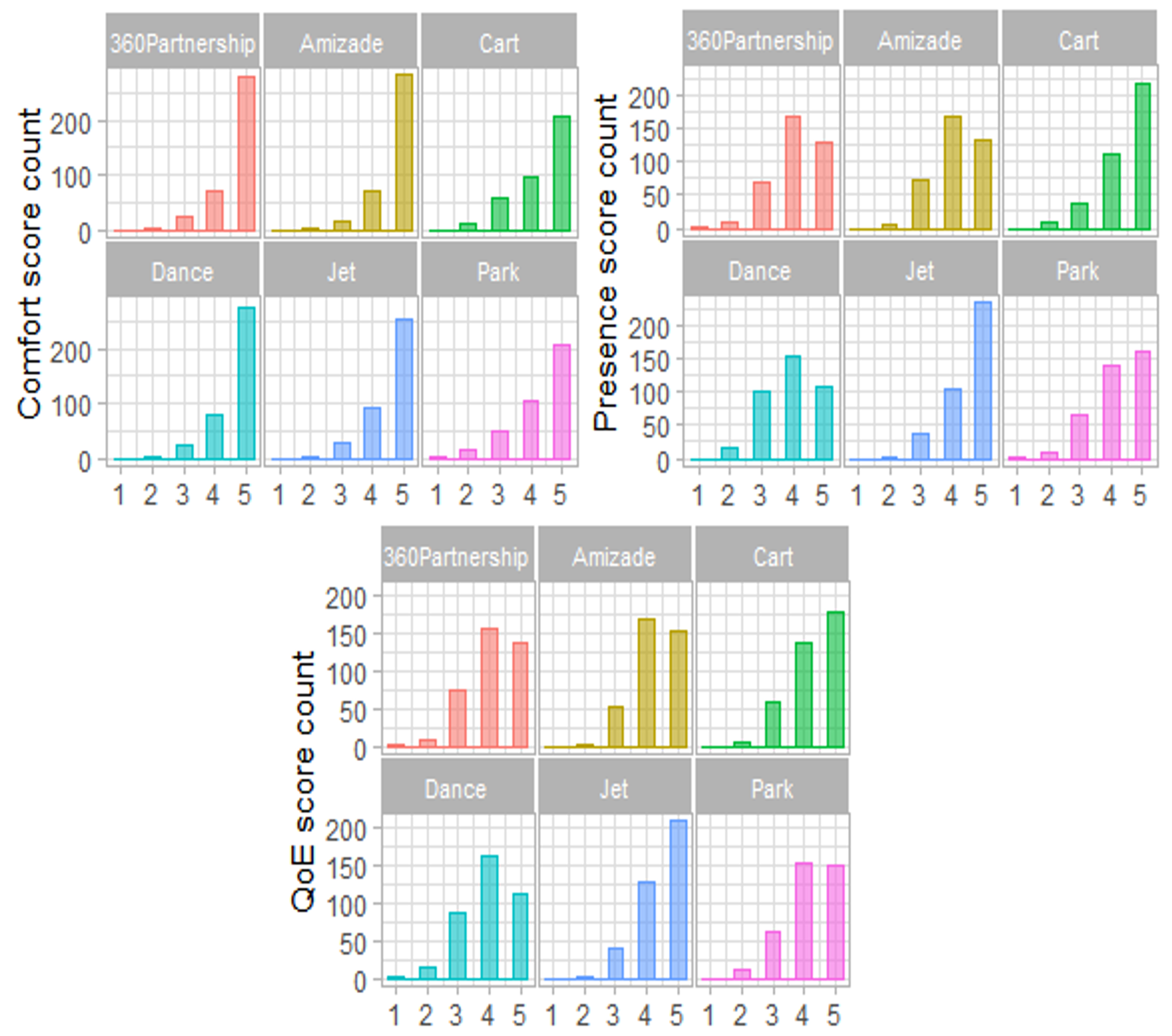}
    \caption{Distributions of presence and comfort scores.}
    \label{fig:distributions}
\end{figure}

First, we use the full dataset to compare the subjective scores under the same content (SRC), evaluating rating distribution and statistical patterns on mean scores. With ratings aggregated by SRC, we plot the histograms in Figure \ref{fig:distributions}, to assess the distribution of rating scores over contents. In terms of comfort, we observe that the scores have a similar trend across all video content. For the ``Cart,’’ ``Park,’’ and ``Jet’’ videos there is a decay in the count of ``excellent'' scores, indicating less comfort for these SRCs. Regarding presence ratings, ``Cart’’ and ``Jet’’ videos distinguish from the others, retaining more ``excellent'' rating counts. In terms of QoE score, ``Dance’’ and ``360Partnership’’ videos follow a similar distribution of presence scores, which indicates a correlation between these factors. Furthermore, by applying a Shapiro test, we confirmed the expected non-normal distribution of subjective ratings $p<0.05$, indicating that inference should consider only non-parametric tests, from now on.  

\begin{figure}
    \centering
    \includegraphics[width=0.8\linewidth]{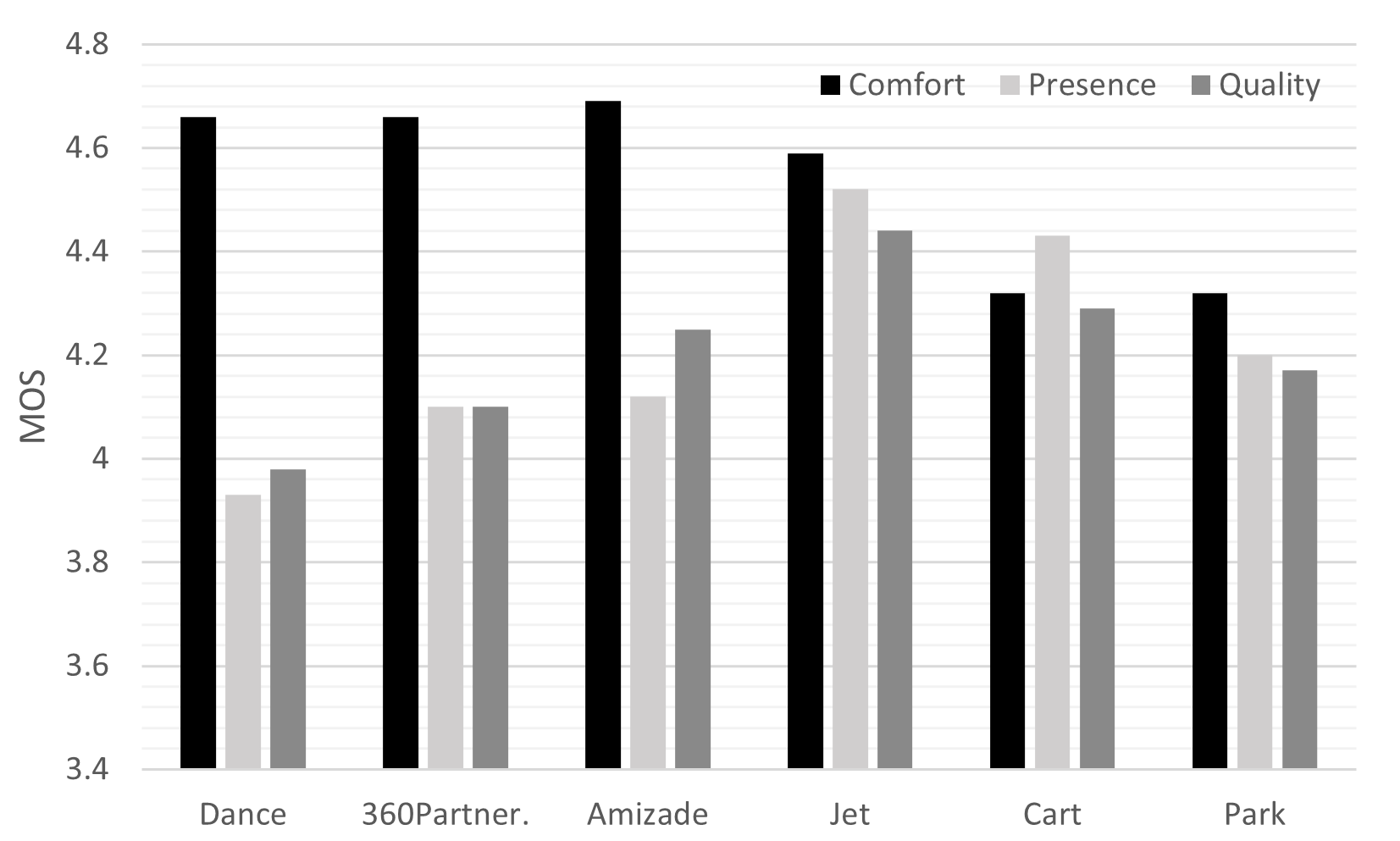}
    \caption{Mean scores for presence, comfort and quality for each video tested.}
    \label{fig:mos}
\end{figure}

Next, we compute equation the MOS (Equation \ref{eq:meanScore}) using all data aggregated by SRC. When looking at the results in Figure \ref{fig:mos}, we did not observe a significant degradation of QoE factor given that all MOS values achieved more than 4, except for the presence and Quality scores of the ``Dance’’ video. This is especially important for the comfort factor since it means users tolerated the wide types of camera motion tested, as well as, in general, they tolerated the alignment edits included in the stimulus. Furthermore, a clear degradation of presence scores was observed for ``Dance’’ video. Some participants reported that this video lacks realism because the dancers in the video seemed to be out of scale, causing estrangement, hence the low $MOS^P$. Other significant feedback provided by participants is that regions of the content resembled watching a conventional video and reduced their feeling of presence, this situation was related to ``Dance'', ``360Partnership,'' and ``Amizade'' videos by different participants. For instance, in ``Amizade'' some participants reported feeling out of the car, and others reported that ``360Partnership'' content appeared artificial because they felt smaller, such situations illustrate how content can cause breaking of the feeling of presence, corroborating with recent studies about realism in VR \cite{Newman2021TheUO}.

Recalling what was observed in Figure \ref{fig:distributions}, three distinguishable groups of MOS were observed. The videos ``Dance'', ``360Partnership'' and ``Amizade'' was characterized by a ``high'' comfort and a low ``presence'' score. Jet had less comfort than those videos, however a higher of presence, and quality score. ``Jet’’ video had the best result when considering all factors followed by ``Cart’’ video, both videos have camera motion, which shows the user's tolerance to motion associated with the alignment edits, later confirmed, the feeling of presence prompt a good overall quality, and that may be related to more engagement with the content \cite{Skola2020VirtualRW}. In agreement with recent studies \cite{Weech2019PresenceAC}, a negative trend between comfort and presence is illustrated in Figure \ref{fig:mos}, while $MOS^P$ increases from $3.93$ (Dance-Fixed) to $4.52$ (Cart-Dynamic) $MOS^C$ decreases from $4.66$ (Dance-Fixed) to $4.32$ (Park-Dynamic). Moreover, considering our camera motion classification (fixed, steady, dynamic), we note a significant slight difference between comfort scores for fixed camera ($MOS^C = 4.66$) and dynamic camera ($MOS^C = 4.32$). In contrast, presence scores have no clear patterns related to camera motion classification.



In Table \ref{tab:correlation}, we show the comparison of subjective scores under the same edit conditions, evaluating whether distinct QoE factors had a statistically significant correlation. From the pooled dataset, we aggregated ratings per edit type, splitting gradual edits into the four rotation speeds. For each subgroup, scores of different QoE factors (presence, comfort, overall quality) were then compared. Initially, we compute the Pearson correlation and the Spearman rank-order correlation coefficients for each subgroup. To interpret correlation coefficients, we follow the convention of \cite{Schober2018CorrelationCA}, where $PLCC,SRCC < 0.1$ is considered negligible. For three edit types (Gradual 10$^\circ$, Instant, and None), there was no correlation between presence and comfort scores, which indicates that these factors have no significant correlation $p<0.05$ in the conditions tested. Furthermore, the same procedure pointed to a significant $p>0.05$, or a weak correlation between scores of comfort and quality $0.1 < PLCC,SRCC < 0.3$, for all edit types. Further, a strong positive correlation between presence and experience $PLCC,SRCC > 0.7$, was found for all edit types. The fact that presence correlates heavily with the overall quality, and comfort correlates moderately, may appear due to the ambiguities of the notion of quality for participants in immersive experiences \cite{itut2021VRexp}. This result points out a similar pattern of QoE factors for FR 10$^\circ$, instant, and none, which indicates that the ratings from those subgroups were indistinguishable. In other words, for low rotation or for instant edits, the users subjectively perceive similarly with that of the original video without alignment edit \cite{dambra2018film}. We then split the dataset by device and observed the same correlation pattern across edit types, however with greater values of correlation between the presence and overall quality $0.66<PLCC<0.74, 0.65<SRCC<0.75$. The positive correlation between comfort and quality scores increases until reach Gradual FR 40$^\circ$, and consistently for all videos we observed that this correlation saturates and tends to decrease for faster rotations (FR 60$^\circ$).









\begin{table}[tb]
    \centering
    \caption{Correlation between QoE factors, with data aggregated by PVS (Edit type). In bold we point the moderate or strong correlation coefficients.}
 \label{tab:correlation}
 \begin{tabular} {c c c c c c}
    \toprule
Edit type&Comparison & PLCC & p-val & SRCC & p-val \\
\midrule
    Gradual &Comf/Pres & 0.021    & 0.681 & 0.021 & 0.688\\
FR 10$^\circ$ &Comf/QoE& \textbf{0.136}       & 0,01& \textbf{0.143} & 0,01\\
&Pres/Exp& \textbf{0.713}     & 0,001& \textbf{0.694}& 0,001\\
    \addlinespace

    Gradual &Comf/Pres& 0.191     & 0,001 & 0.142& 0,001\\
FR 20$^\circ$&Comf/QoE & \textbf{0.339}     & 0,001 & \textbf{0.319}& 0,001\\
&Pres/Exp& \textbf{0.720}     &  0,001& \textbf{0.709}& 0,001\\
    \addlinespace

    Gradual  &Comf/Pres & 0.205     & 0.059& 0.175& 0,001\\
FR 40$^\circ$&Comf/QoE& \textbf{0.396}      &  0,001& \textbf{0.360}& 0,001\\
&Pres/QoE& \textbf{0.732}     & 0,001& \textbf{0.736}& 0,001\\
    \addlinespace

    Gradual &Comf/Pres& 0.173     &  0,001& 0.132& 0.01\\
FR 60$^\circ$&Comf/QoE& \textbf{0.363}     &  0,001& \textbf{0.328}&  0,001\\
&Pres/Exp& \textbf{0.716}     &  0,001& \textbf{0.710}&  0,001\\
    \addlinespace

    Instant&Comf/Pres& 0.045    & 0.385& 0.028& 0.591\\
&Comf/QoE& 0.150     &  0,01& 0.157&  0,01\\
&Pres/QoE& \textbf{0.674}     &  0,001&\textbf{0.658} & 0,001\\
    \addlinespace

    None &Comf/Pres& 0.001   & 0.988& 0.007& 0.887\\
&Comf/QoE& 0.236     &  0,01& 0.225& 0,01\\
&Pres/QoE& \textbf{0.672}     &  0,001& \textbf{0.675}&0,001\\
\hline
\end{tabular}
\end{table}

For comparative analysis, we will be restricted to the subjective factors of comfort and presence, that had negligible or low correlation. Following, we test the impacts of video content on the presence and comfort assessment. We performed the non-parametric Kruskall-Wallis H test (KW) since Levene's test did not confirm the assumption of normality and homogeneity of the variance, for both factors.  First, we conducted a KW test with the full dataset, a statistically significant effect of content on presence ($\chi^2 = 20.376, df=4,p=<0.001$) and on comfort $(\chi^2 = 31.423, df=4,p=<0.001)$ was observed. To go into detail, we conducted a post hoc KW test with a false discovery rate (FDR) p-value adjustment. For this post hoc test, we define a pair of SRC to compare, aggregate all scores of comfort and compare between SRC-grouped scores. After performing it for all pairs of SRC, we apply the same test method for presence. We present the results of this analysis in Table \ref{tab:PairTest_content}. In terms of comfort, no significant differences were found when comparing the distributions of videos inside the same category, i.e., Dance-360Partnership, Amizade-Jet, Cart-Park. Meaning that the camera dynamic categorization successfully fitted the content. Extracting these inter-class comparisons, from the 12 remaining content pairs, we found the odds of 7:12 significant difference $p<0.05$ for presence, and 8:12 for comfort.
Furthermore, by clustering video pairs with no significant difference in presence, we found two clusters, videos ``360Partnership'', ``Amizade'', ``Park'' and ``Dance'' are in one group, while ``Jet'' and ``Cart'' are on the other. These clusters substantiate our previous observations, from score distribution Figure \ref{fig:distributions}. A specific feature ``Jet'' and ``Cart'' have in common, differently from other videos is that they have an atmosphere of action or challenge, but it is difficult to determine a single feature that causes the sense of presence, for example, the sequence ``Amizade'' also action and the video was recorded in the city of the participants, what could be also related to more engagement and in consequence greater presence scores.

\begin{table}[tb]
    \centering
    \setlength{\tabcolsep}{4pt}
    \caption{Paired Kruskall-Wallis test with FDR adjusted p-values for presence and comfort scores.}
 \label{tab:PairTest_content}
 \begin{tabular} {c c c l l}
    \toprule
    Camera &Video & Comparison &\multicolumn{2}{c}{\textbf{p-value}} \\
    Dynamic&&Video & \small{Comf.} & \small{Pres.} \\
    \midrule
    Fixed & Dance & \footnotesize{360Partnership} & \multicolumn{1}{c}{1} & \multicolumn{1}{c}{0.748}\\
    &&  Amizade & \multicolumn{1}{c}{1} & \multicolumn{1}{c}{0.711}\\
    &&Jet & \multicolumn{1}{c}{1} & \textbf{$\scriptscriptstyle<$ 0.001}\\
    \addlinespace
    & \footnotesize{360Partnership} &  Amizade & \multicolumn{1}{c}{0.184} & \multicolumn{1}{c}{0.914}\\
    &&Jet & \multicolumn{1}{c}{0.817}& \textbf{$\scriptscriptstyle<$ 0.001}\\
    \hline
    Steady&Amizade &  Jet & \multicolumn{1}{c}{0.108}& \textbf{$\scriptscriptstyle<$ 0.001}\\
    &&Cart & \textbf{$\scriptscriptstyle<$ 0.001}& \textbf{$\scriptscriptstyle<$ 0.001}\\
    &&Park & \textbf{$\scriptscriptstyle<$ 0.001}& \multicolumn{1}{c}{0.712}\\
    \addlinespace
    
    &Jet &  Cart & \textbf{$\scriptscriptstyle<$ 0.05}& \multicolumn{1}{c}{0.968}\\
    &&Park & \textbf{$\scriptscriptstyle<$ 0.01}& \textbf{$\scriptscriptstyle<$ 0.01}\\
    \hline
    Dynamic&Cart & Park & \multicolumn{1}{c}{0.667} & \textbf{$\scriptscriptstyle<$ 0.01}\\
    &&Dance & \textbf{$\scriptscriptstyle<$ 0.01} &\textbf{$\scriptscriptstyle<$ 0.001}\\
    &&\footnotesize{360Partnership} & \textbf{$\scriptscriptstyle<$ 0.01}& \textbf{$\scriptscriptstyle<$ 0.001}\\
    \addlinespace

    &Park & Dance & \textbf{$\scriptscriptstyle<$ 0.001}& \multicolumn{1}{c}{0.377}\\
    &&\footnotesize{360Partnership} & \textbf{$\scriptscriptstyle<$ 0.01}& \multicolumn{1}{c}{0.677}\\
   \hline   
 \end{tabular}
 \end{table}

\subsection{Comparison between instant and gradual edit}


\begin{figure}[htb]
    \centering
    \includegraphics[width = 1.\linewidth]{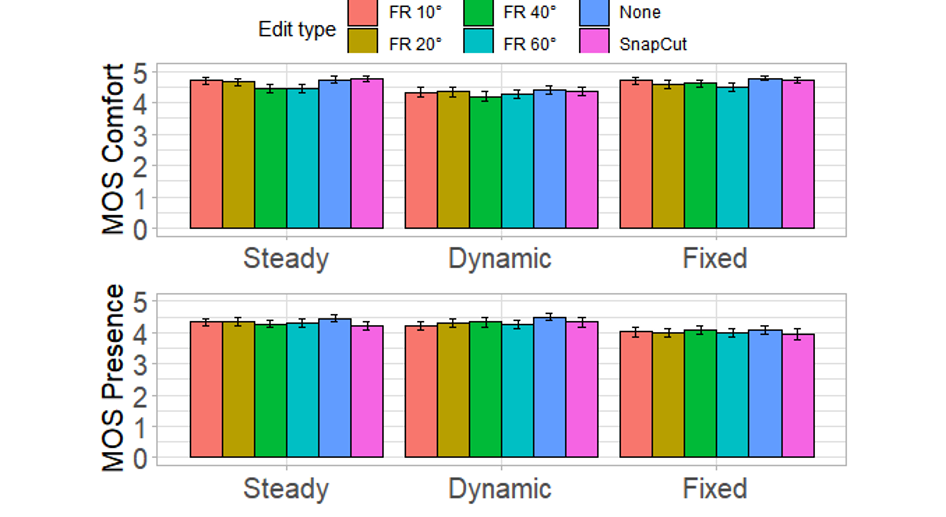}
    \caption{Subjective responses according to editing type and gradual edit rotation speeds.}
    \label{fig:InteractionPerception}
\end{figure}

\begin{figure}
    \centering
    \includegraphics[width = 0.9\linewidth]{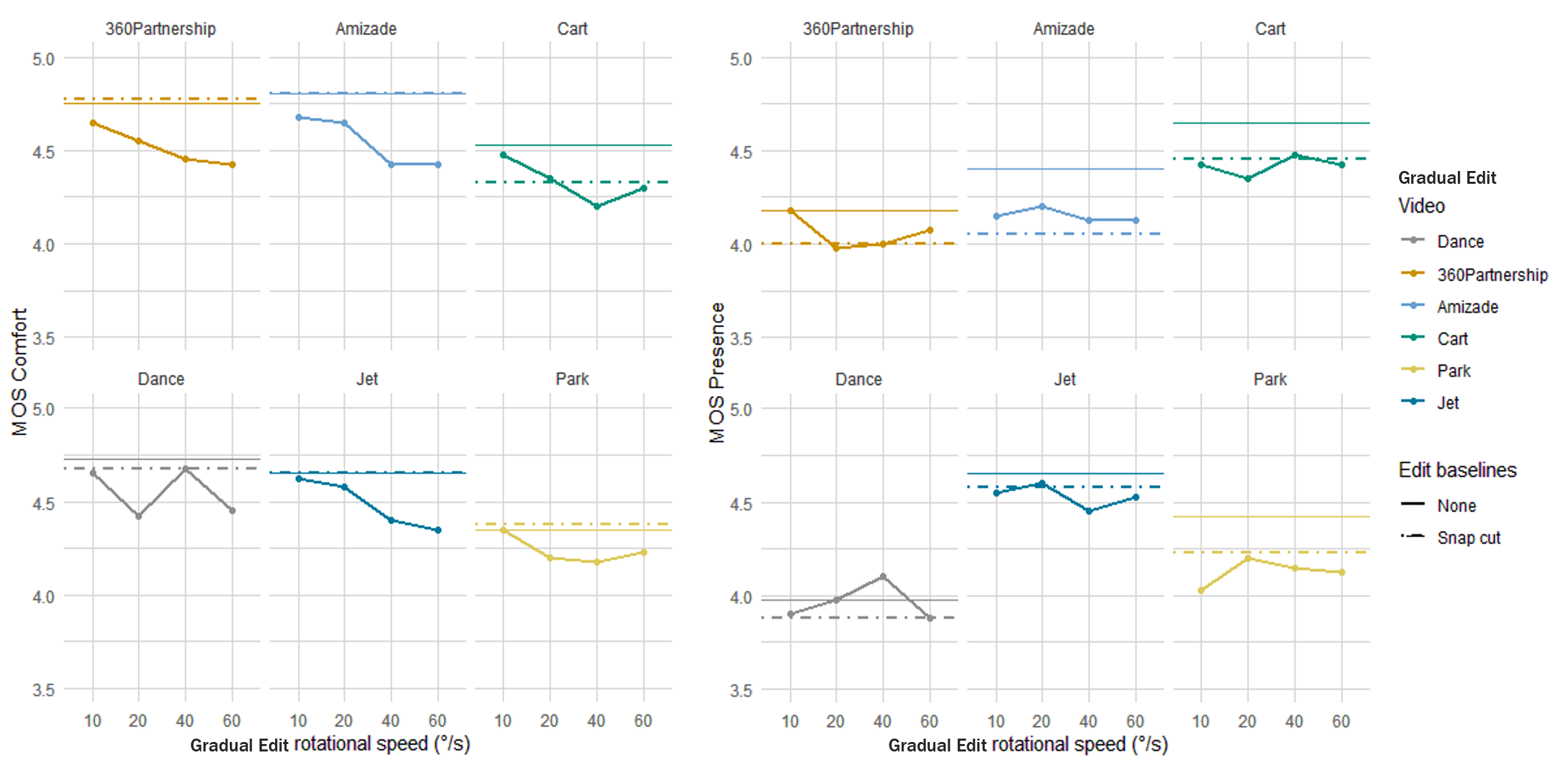}
    \caption{Baseline plot for MOS of presence and comfort. Comparing all rotation speed versions of our gradual edit, called fade-rotation (FR), against the instant edit (snap-cut) and no edit (none) for each video.}
    \label{fig:pres_baseline}
\end{figure}

In Figure \ref{fig:InteractionPerception}, we observe the results of the mean opinion score (MOS) of presence, and comfort for all tested edit types with content grouped in terms of scene dynamic category. In terms of comfort, we note a tendency for comfort to decay depending on the rotation speed of the gradual (FR) edit. To test this tendency, we filtered the dataset with only the FR edits and tested for the effect of comfort score and the edit type by applying a KW test. We found a significant effect ($\chi^2 = 12.511, df=3,p<0.01$) for comfort, and a not significant effect for presence ($\chi^2 = 0.236, df=3,p>0.05$).
By filtering scores into two groups, each filtered with samples from a single edit type, we test their statistical difference by applying Welch's two-sample t-test with FDR correction. First considering comfort and the hypothesis \textbf{H1}, with the full dataset, we found a significant difference between snap-cut and FR 40$^\circ$, between snap-cut and FR 60$^\circ$ $p<0.001$. Filtering data by camera motion and selecting only dynamic type, we found a significant difference between snap-cut and all types of gradual edit $p<0.001$. Meaning a strong indication that for some types of content motion, the edit type influences the comfort of the users. In the dynamic motion filter, all FR edit had equivalent 95\% confidence MOS of that of the snap-cut edit ($MOS^C(sc) = 4.36 \pm 0.08, MOS^C(FR10^{\circ}) = 4.34 \pm 0.07 , MOS^C(FR20^{\circ}) = 4.35 \pm 0.08, MOS^C(FR40^{\circ}) = 4.20 \pm 0.08, MOS^C(FR60^{\circ}) = 4.27 \pm 0.07$). For fixed, only one FR edit had equivalent 95\% confidence MOS ($MOS^C(sc) = 4.74 \pm 0.04, MOS^C(FR10^{\circ}) = 4.71 \pm 0.05$). For steady motion only two FR edit ($MOS^C(sc) = 4.77 \pm 0.05, MOS^C(FR10^{\circ}) = 4.71 \pm 0.05, MOS^C(FR20^{\circ}) = 4.67 \pm 0.05$). That said, we were able to accept \textbf{H1} for the FR10$^{\circ}$ in all camera motion types. For dynamic motion content, \textbf{H1} can be accepted for all gradual edit types. In the case of steady camera motion, we accepted \textbf{H1} for FR10 and FR20. In conclusion, FR20$^{\circ}$, FR40$^{\circ}$ FR60$^{\circ}$ may be avoided to be implemented in a video player that does not account for camera motion. FR10$^{\circ}$ or snap-cut is preferable since they have a lower probability to imply discomfort, in the conditions tested. Figure \ref{fig:pres_baseline} illustrate baseline comparison for data grouped by video contents, we found again that in all videos FR10$^{\circ}$ gradual edit had equivalent 95\% confidence MOS of snap-cut, and no significant difference $p>0.01$. As a side case, we observed that for videos ``Cart'' and ``Dance'' all gradual edits had no significant difference with snap-cut. In these videos, snap-cut had the worst MOS. From qualitative feedback, we observed that in the case of ``Cart', the instant edit was uncomfortable because was combined with the strong camera translation '. In the case of ``Dance,'' two qualitative feed backs pointed that video instant cut interrupted the shift between dance groups. It is clear from Figure \ref{fig:pres_baseline} that comfort tends to decrease with the rotation speed, however, specific conditions can break this trend. For instance, the video ``Dance’’ surprisingly had the best results for FR40$^{\circ}$, presence and quality scores followed the same peak. Showing that there is a non-trivial relation between FR rotation speed and content.

As noted from Figure\ref{fig:InteractionPerception}, when compared to comfort, presence has flatter MOS across all edit types. The slower versions of FR (FR10$^{\circ}$,20$^{\circ}$) did not have better MOS as in comfort scores, even the FR40$^{\circ}$ had the higher MOS for dynamic and fixed camera motions. We show in Figure\ref{fig:InteractionPerception} the results grouped by content. Considering presence scores across content, we observe a similar trend for all edit types. However, we point out for specific cases where gradual had higher 95\% confidence result, for example in the video ``Dance'' FR40$^{\circ}$ worked even better than the ``None'' baseline, and FR10$^{\circ}$ ``360Partnership''. Those two videos had fixed cameras and indoor scenes. Moreover, the presence scores were low for all edit types, gradual rotation had good results. However, it is not clear what in the content could be the source of that good result for gradual edits. Also, in the video ``Amizade'', a steady camera motion scenario, gradual rotations had slightly (but with $p>0.05$) better MOS than instant edit. Considering \textbf{H2} hypothesis, we applied a Welch's t-test in the full dataset and did not find any statistically significant difference between edit types. Showing that edit type has no significant effect on viewers' presence. Given the precedent tests, we rejected the hypothesis \textbf{H2}, confirming that, in the case tested, snap-cut and fade-rotation had no distinguishable effect for presence.

\begin{figure}
    \centering
    \includegraphics[width = .6\linewidth]{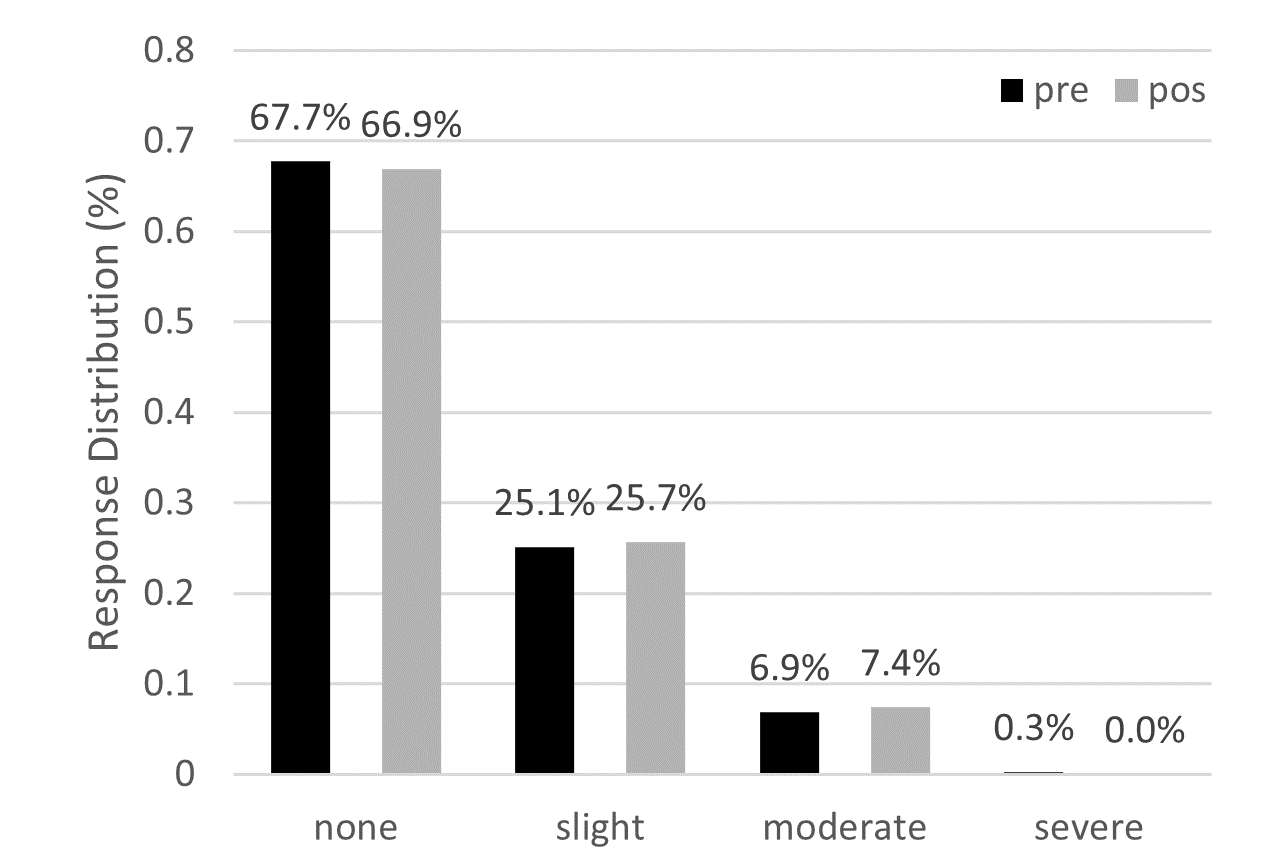}
    \caption{Distribution of all cybersickness symptoms}
    \label{fig:ssq}
\end{figure}

Figure \ref{fig:ssq} shows the overall cybersickness symptoms count, from the applied SSQ before and after. We observe almost the same results for pre and post. More than 90\% of the were reported as none or slight. Only one severe symptom was observed, this happened in the ``Jet’’ video, with one participant that had height phobia. It is known that individual sensibilities can provoke unequal effects in VR \cite{Howard2021AMO}. 

\subsection{Analysis of Head Motion Metrics}
The effects of alignment in viewers comfort and presence was measured by two metrics: first, the alignment after edit $A$ with data grouped by ``non-aligned'' and ``aligned'' observations. Second, the alignment state before and after the edit (see \ref{sec:HM_metrics} for details). To test \textbf{H3}, this is, if alignment performance has an impact in subjective assessment scores, we compared the set of scores from the two groups. We performed 15 Wilcoxon rank sum tests with continuity correction, since there are five edit types (none was not considered) and three factors tested. Just one comparison got a $p<0.05$, for FR 10$^\circ$ in overall quality score. This is a strong indication that the alignment did not have impact on subjective assessment answering. 

Since after filtering we had unbalanced data, we applied analysis of variance considering $A$ as treatment alone, and again no significant difference was observed. Following, we performed a TukeyHSD post hoc to account for $A$ and content (21 comparisons), $A$ and Edit type (15 comparisons), as well as $A$ and camera motion type (6 comparisons), neither alone nor the interacting factor had significant differences, in total we performed 42 non significant comparisons. Concluding, no statistical distinguishable difference was found between group ``non-aligned'' and group ``aligned''. With that we rejected hypothesis \textbf{H3}. 

\begin{figure}
    \centering
    \includegraphics[width = 0.9\linewidth]{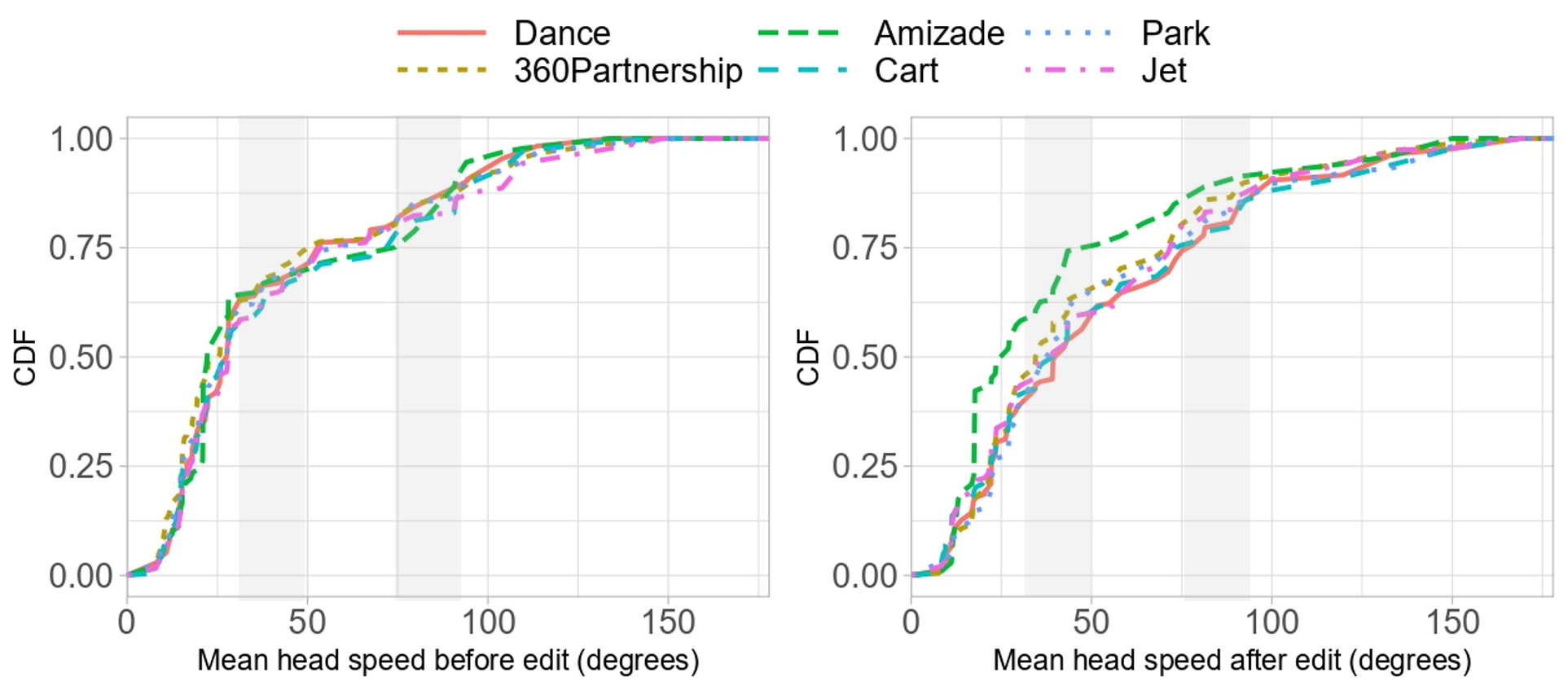}    
    \caption{CDF of the head speeds aggregated by video content. Left: CDF before the edits Right: after the edit. Intervals where resides two quarters (25\%-75\%) was highlighted.}
    \label{fig:cdf}
\end{figure}

\begin{figure}
    \centering
    \includegraphics[width = 1.\linewidth]{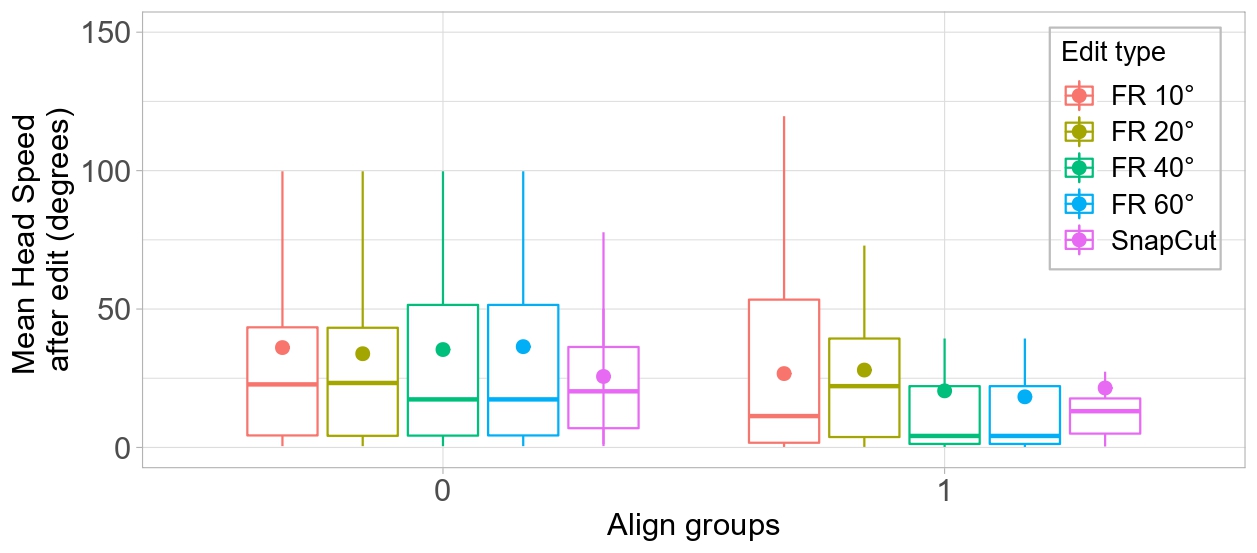}
    \caption{Boxplots of the \textit{mean head speed after edit} for each edit type. Anova test showed significant difference ($<0.001$) between both aligned and non-aligned groups. The circles shows the mean after outliers removal.}
    \label{fig:SpeedBoxplot}
\end{figure}

Finally, we should test \textbf{H4} related to head motion reduction. Figure \ref{fig:cdf} illustrates the mean head speed CDF before and after alignment edits, except for the video ``Amizade'' the plateu of the CDF goes down referring to broader distributions of speeds, whereas for 75\% of the speed distribution almost all videos had head speed less or equal to 50$^\circ$/s, while after edit speed less or equal to 75$^\circ$/s was observed. We performed an Anova Omnibus test between aligned and non-aligned groups and found a significant reduction in head speed after edit $F(30.95,1, p<001)$. Figure \ref{fig:SpeedBoxplot} shows the boxplot of the mean speed per edit type in two groups, the ``aligned'' (1) and ``non-aligned'' (0). For the aligned group we observe that FR 40$^\circ$ has the lower mean: 8\%  lower than snapcut, followed by FR 60$^\circ$ with 2.46\% lower than snap-cut, the total mean head speed reduction was FR 10$^\circ$ = 14.9$^\circ$, FR 20$^\circ$ = 9.5, FR 40$^\circ$ = 26.7$^\circ$, FR 60$^\circ$ =33.1$^\circ$, Snap-cut = 21.5$^\circ$. A common consequence between all edit types tested is that they affected reducing head speed, which may be related to the fixation at ROI, reducing the exploratory behavior, this stability effect agrees with the literature \cite{sassatelli2019user} and can be useful for streaming applications. As a remark, for this analysis we filtering head speed higher than 150$^\circ$/s (refer to Section \ref{sec:hm} for description). With this observations we confirm \textbf{H4}.

\section{Final remarks}
\label{sec:conclusion}

From the goals established, we could conclude that: 

\begin{enumerate}
    \item all edit types had high scores for presence, and it was not possible to discriminate edit types statistically;
    \item gradual edit (fade-rotation) is acceptable in terms of comfort, for a maximum angular speed of 20°, where the criteria of acceptability were being at the same level as the instant edit;
    \item content was found to play a significant role in comfort and presence and device did not have a significant impact on subjective assessment. Specifically, for content where there are accelerated elements relative to the camera (or that the camera moves with prominent acceleration) presence tends to increase, and comfort the opposite;
    \item we found that the alignment performance does not impact the subjective assessment of presence and comfort;
    \item the alignment performance impacted the behavior right after the edit, specifically the head motion 1 s after the edit was reduced, resulting in a result 8\% lower than that of the instant edit.
\end{enumerate}

To sum up, we conclude that it is clear that gradual edit (fade-rotation) for slower rotation speeds (FR10$^{\circ}$, FR20$^{\circ}$) can not be considered a worst solution than instant edits. Considering these fade-rotation versions, that was acceptable in terms of comfort, we recommend an angular rotation speed of 10°, which was able to reduce head speed in 14.9°/s.

Furthermore, our results give insight into how adaptive alignment edits affect subjective QoE and how to measure their impacts on head motion metrics. To implement gradual edit, please follow the parameters described in \ref{sec:stimuli-parameter}. We were able to confirm that fade-rotation is viable and useful as a transition technique for specific cases. A dynamic version of the edit strategies, where edits are implemented in playback time, taking into account the actual head motion data to adapt the edit. Therefore, we place the study of a dynamic version of such edit strategies as complementary future work. Together with studies of similar nature, results from our study are only rigorously valid for our chosen stimuli. We chose to deal with short clips, even though extracted from real videos, to be able to conduct a multifactorial analysis controlling the impact of stimuli length and other confounding factors. For that reason, some outcomes may not be generalized to a realistic use case. Another limitation of our study is that we did not include a pure rotation edit as a baseline condition; this restricted our analysis because we could not measure the benefits of including the fade-in-out effect.

\bibliographystyle{IEEEtran}
\bibliography{bib}

\end{document}